\documentclass[11pt,twoside]{article}

\usepackage{asp2006}
\usepackage{epsf}
\usepackage{lscape}

\begin{document}
\title{A Multi-Scale Study of IR and Radio Emission 
from M33}   
\author{F.  S. Tabatabaei\altaffilmark{1}, R. Beck\altaffilmark{1}, M. Krause\altaffilmark{1}, E. M. Berkhuijsen\altaffilmark{1}, R. Gehrz\altaffilmark{2}, K. D. Gordon\altaffilmark{3}, J. L. Hinz\altaffilmark{2}, and G. H. Rieke\altaffilmark{2}}   
\altaffiltext{1}{Max-Planck Institut f\"ur Radioastronomie, Auf dem H\"ugel 69, 53121 Bonn, Germany}
\altaffiltext{2}{Steward Observatory, University of Arizona, 933 North Cherry Avenue, Tucson, AZ 85721}
\altaffiltext{3}{Space Telescope Science Institute, Baltimore, MD 21218}
\begin{abstract} 
The  origin  of  the  tight  radio--IR  correlation in  galaxies  has   not   been    fully   understood.   One  reason   is  the  uncertainty about  which  heating sources  (stars or diffuse interstellar  radiation  field)   provide   the   energy   that  is absorbed by dust  and  re-radiated in IR.  Another problem   is  caused  by  comparing the IR emission with the thermal  and   nonthermal   components  of   the   radio   continuum emission   separated     by    simplistically     assuming    a  constant      nonthermal      spectral     index.  We use the data at the $ Spitzer$ MIPS wavelengths of 24, 70, and 160\,$\mu$m, as well as  recent radio continuum map at 3.6\,cm observed with the 100--m Effelsberg telescope.  Using    the   wavelet   transformation,  we   separate  diffuse   emission  components from  compact sources and  study  the  radio-IR  correlation  at  various  scales. We also investigate the IR correlations with the thermal and nonthermal radio emissions separated by our developed method. A H$\alpha$  map serves  as  a tracer of  star forming regions.
\end{abstract}


\section{Wavelet Analysis of MIPS}
The  2D–-wavelet  transformation is a useful tool  to separate diffuse emission from that of compact sources (e.g. Frick et al. 2001). A comparison of the wavelet-decomposed images of  MIPS (Hinz et al. 2004) indicates that the 24 and 70\,$\mu$m emission emerge mostly from the compact structures corresponding to star forming and HII regions. The 160\,$\mu$m emission emerges from both compact and extended structures. The 160\,$\mu$m wavelet spectrum (Tabatabaei et al. 2005, 2007a) shows an increase when it reaches the scale of complexes of dust and gas clouds of $\sim$\,1 kpc, and a second increase in transition to the large-scale structures of diffuse dust emission. 
  
\section{ Radio ($\lambda$3.6 cm)- FIR and H$\alpha$ Correlations}   
The 3.6\,cm emission from M33 (Tabatabaei et al., 2007b) shows better correlations  
with  warm  than  with cold   dust  at scales smaller than the width of  the spiral   arms  (1.6 kpc).  The  results were     compared   before  and   after  subtracting the 11 brightest  HII  regions.  Comparing with the H$\alpha$ map, the role of star forming regions in heating  the dust, even the cold   dust,   is   generally    important    at   scales   up  to  4 kpc.


After  separating the thermal and nonthermal components of the radio continuum emission (using the de-extincted H$\alpha$ map as a thermal emission template, Tabatabaei et al. 2007c), not only the stronger  warm dust--thermal radio than the cold dust--nonthermal radio correlation is proved, but also a stronger warm dust--nonthermal radio than the cold dust--nonthermal radio correlation is found (Fig.~1). It is also found that the cold dust has a better correlation with the thermal than with the nonthermal emission. The  IR--nonthermal  radio correlation  is  better at  scales    between  0.8 and 2\,kpc, scales  including giant molecular clouds,  spiral   arms,  and the central   extended   region  of  M33.
\begin{figure}
\begin{center}
\resizebox{\hsize}{!}{\includegraphics*{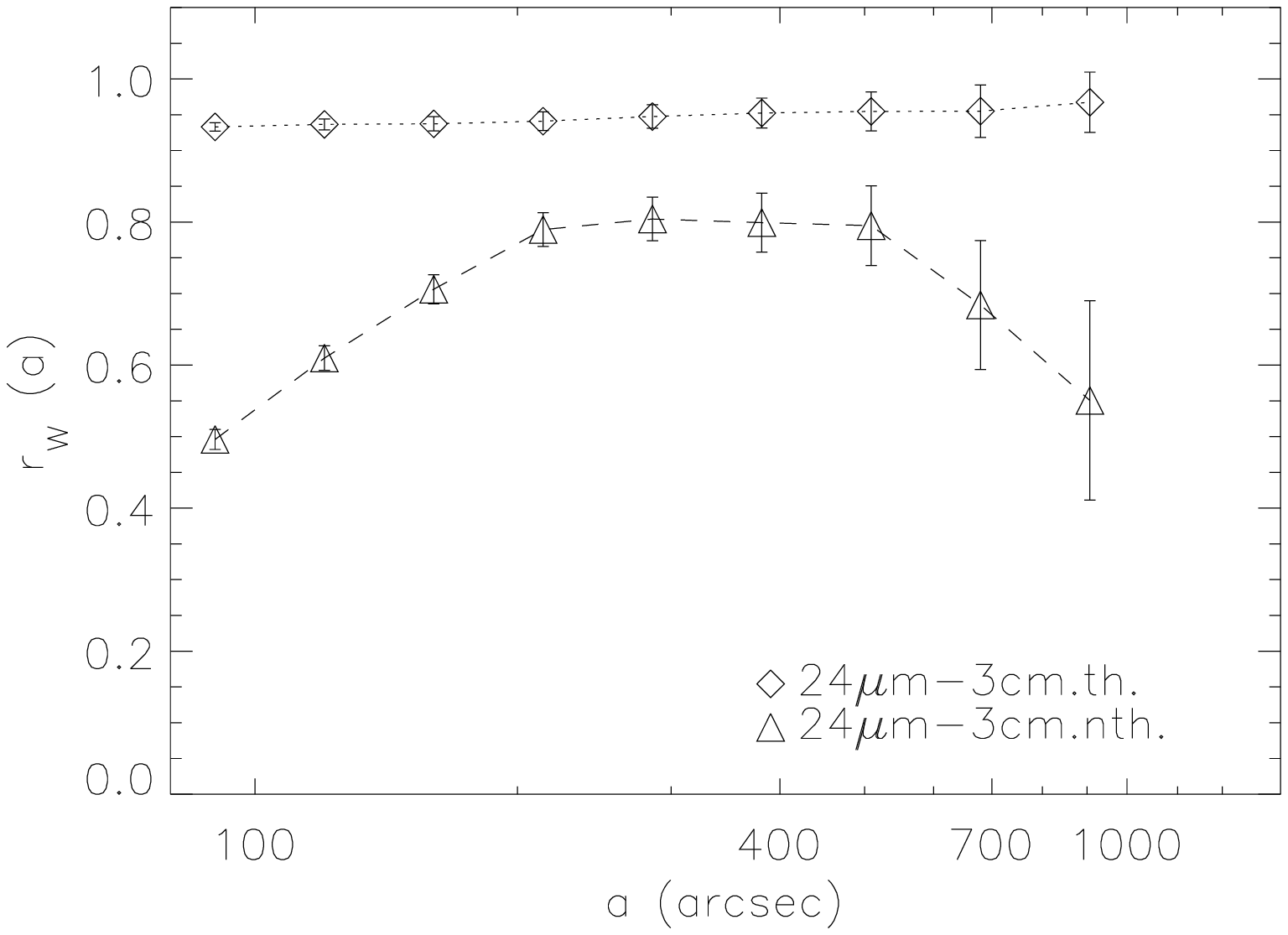}
\includegraphics*{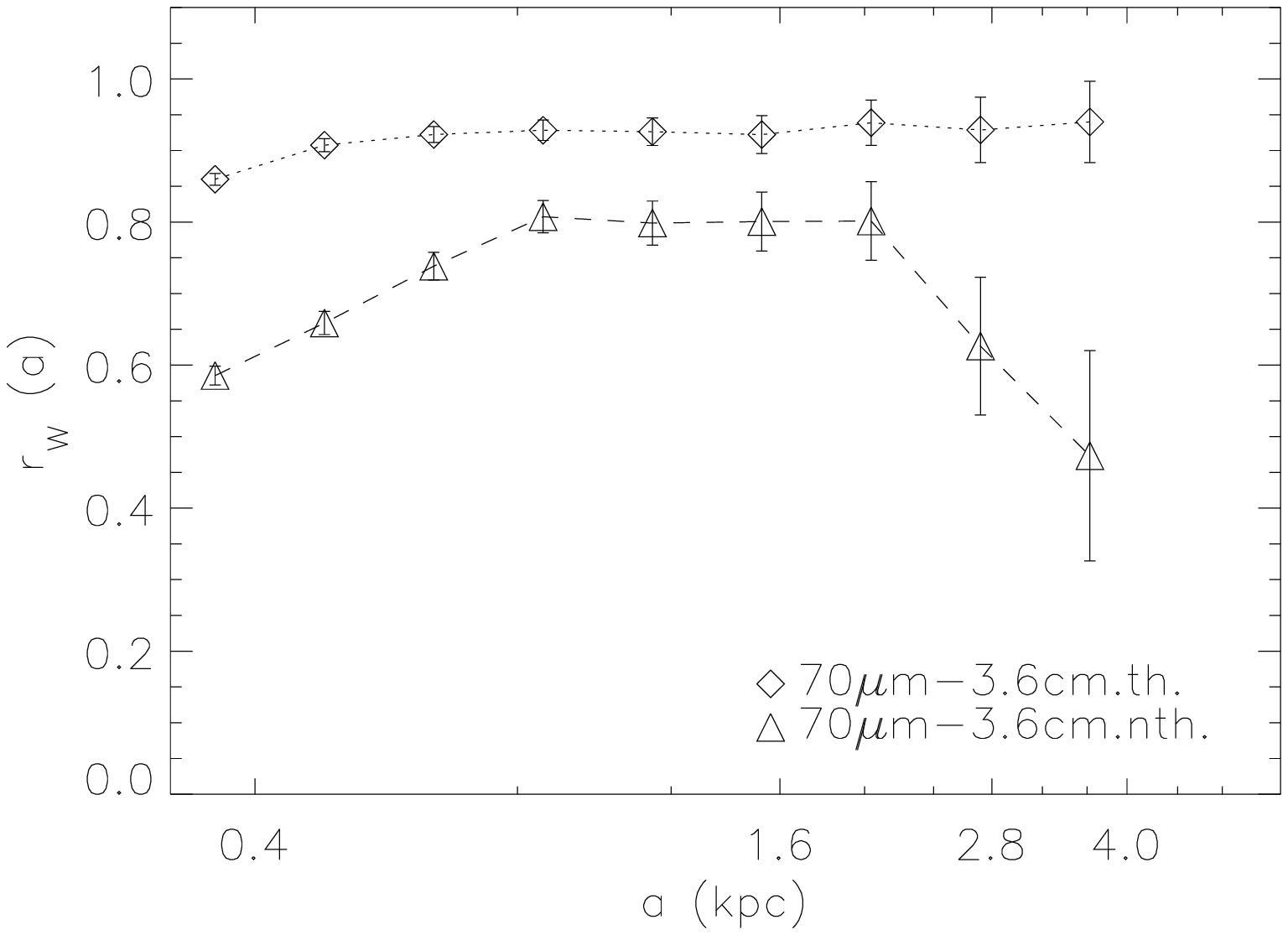}
\includegraphics*{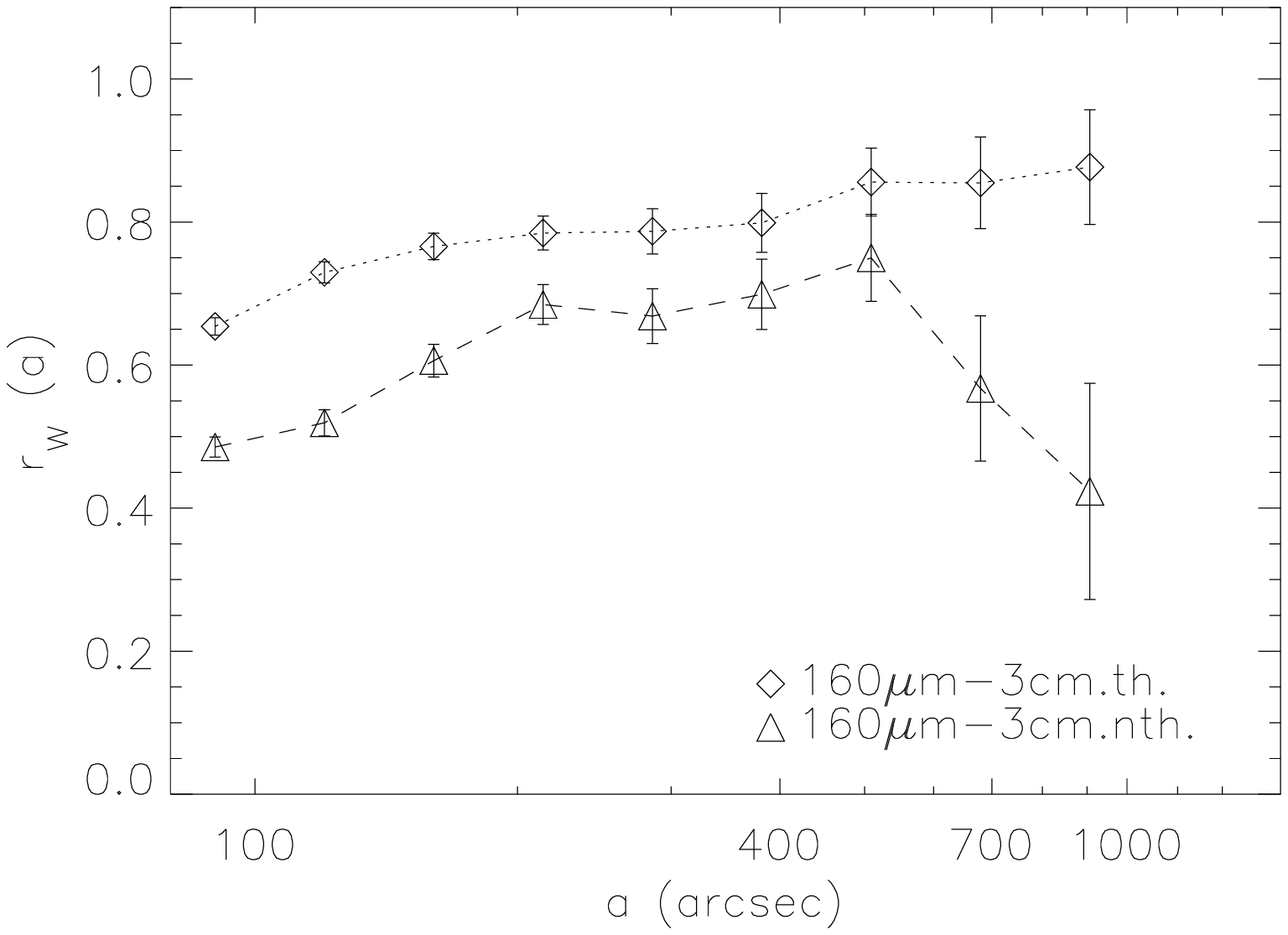}}
\caption[]{The cross--correlation between the thermal/nonthermal components of the 3.6\,cm  and 24\,$\mu$m (left), 70$\mu$m (middle), and 160\,$\mu$m (right) maps. }
\end{center}
\end{figure}

\section{Conclusions}
- HII   regions   influence  the  IR   emission  with a    strength  inversely depending on wavelength: more influence at  24 $\mu$m  and   less influence at 160\,$\mu$m.\\
- The  nonthermal  radio   emission   correlates  well with  the  thermal radio and H$\alpha$ emission out to the  scale of the central extended region, 2.5\,kpc.\\
- The warm dust--thermal radio correlation  is  much  stronger than the cold  dust--nonthermal  radio  correlation at   scales  smaller  than 4 kpc.\\
- At these scales,  the role of UV photons  from  O/B  stars  in heating    the  cold   dust  is  more  significant than  of  other  energy sources.\\
- The  better  correlation  of the nonthermal emission with the 24\,$\mu$m  than  with  the 160\,$\mu$m indicates  that  magnetic field is stronger in the star forming regions. However, there is no indication that the magnetic field is enhanced at scales of the HII complexes. 




\begin{thebibliography}{}
\bibitem[]{}Frick, P., Beck, R., Berkhuijsen, E.~M., \& Patrickeyev, I.~T., 2001, MNRAS, 327, 1145
\bibitem[]{}Hinz, J.~L., Rieke, G.~H., Gordon, K.~D., et al. 2004, ApJ, 154, 259
\bibitem[]{}Tabatabaei, F., Krause, M., \& Beck, R. 2005, Astronomische Nachrichten, 326, 532 
\bibitem[]{}Tabatabaei, F.~S., Beck, R.,  Krause, M., et al. 2007a, A\&A, 466, 509
\bibitem[]{}Tabatabaei, F.~S., Krause, M., \& Beck, R. 2007b, A\&A, 472, 785 
\bibitem[]{}Tabatabaei, F.~S., Beck, R., Kr\"ugel, E., et al. 2007c, A\&A, 475, 133

\end{thebibliography}
\end{document}